\documentclass[column,showpacs,preprintnumbers,amsmath,amssymb]{revtex4}
\preprint{}
\usepackage{graphicx}
\usepackage{dcolumn}
\usepackage{bm}
\usepackage{mathrsfs}
\begin{document}
\title{Enhancing precision of damping rate by PT symmetric Hamiltonian}
\author{Dong  Xie}
\email{xiedong@mail.ustc.edu.cn}
\affiliation{College of Science, Guilin University of Aerospace Technology, Guilin, Guangxi, P.R. China.}

\author{Chunling Xu}
\affiliation{Faculty of Science, Guilin University of Aerospace Technology, Guilin, Guangxi, P.R. China.}

\begin{abstract}
We utilize quantum Fisher information to investigate the damping parameter precision of a dissipative qubit. PT symmetric non-Hermitian Hamiltonian is used to enhance the parameter precision in two models: one is direct PT symmetric quantum feedback; the other is that the damping rate is encoded into a effective PT symmetric non-Hermitian Hamiltonian conditioned on the absence of decay events. We find that compared with the case without feedback and with Hermitian quantum feedback, direct PT symmetric non-Hermitan quantum feedback can obtain better precision of damping rate. And in the second model the result shows that the uncertainty of damping rate can be close to 0 at the exceptional point. We also obtain that non-maximal multiparticle entanglement can improve the precision to reach Heisenberg limit.
\end{abstract}

\pacs{03.65.Yz; 03.65.Ud; 42.50.Pq}
\maketitle

\section{Introduction}
Quantum metrology is becoming a more and more important subject, which
concerns the estimation of parameter precision under the constraints of
quantum mechanics \cite{lab1,lab2,lab3}. There are widespread applications such as in timing, healthcare, defence, navigation,
 astronomy and magnetometry\cite{lab4,lab5,lab6,lab7,lab8,lab9}. For the mean-square error criterion, the quantum Cram$\acute{e}$r-Rao
bound\cite{lab10,lab11,lab12} is the most known analytic bound
which shows that the precision of the parameter is inversely proportional with quantum Fisher information(QFI). Namely, QFI plays a central role in quantum metrology. And QFI also connects with other quantities, such as, non-Markovianity\cite{lab13}, quantum phase transition\cite{lab14}.

Quantum system inevitably interacts with its environment, which induces decoherence. Generally, decoherence deceases the precision of many unitary parameters, such as, frequency\cite{lab15,lab16} and phase\cite{lab161}.
In order to enhance the precision of unitary parameters, there are a lot of various methods, such as quantum error
correction, dynamical decoupling, decoherence-free subspace, and reservoir engineering\cite{lab17,lab18,lab19,lab20,lab21}, proposed to suppress the decoherence.
While unitary transformations have occupied most of the attention in the realm of quantum metrology, the full characterization of a system would also require
the estimation of decoherence parameters. In ref.\cite{lab22}, the simultaneous estimation of phase and dephasing for qubits was proposed. In ref.\cite{lab23}, temperature was measured by estimating the dephasing parameter.

Recently, Qiang Zheng et.al\cite{lab24} suggested an alternative method, direct quantum feedback, to enhance the damping parameter precision of optimal quantum estimation of a dissipative qubit. In this reference, the feedback process is dominated by Hermitian Hamiltonian. And there are a lot of works\cite{lab25,lab26,lab27,lab28} bout quantum feedback dynamics which depend on Hermitian Hamiltonian. In this article, we utilize PT symmetric non-Hermitian feedback Hamiltonian to enchance the damping parameter precision. As a result,  compared with the case without feedback, direct PT (parity and time) symmetric non-Hermitan quantum feedback can obtain better precision of damping rate. To take into account the time for measuring damping rate, direct PT symmetric non-Hermitian quantum feedback can obtain better precision of damping rate than the case of Hermitian feedback.

In addition, conditioned on the absence of decay events, the damping rate is encoded into an effective PT symmetric non-Hermitian Hamiltonian\cite{lab29,lab30}. We find that the uncertainty of damping rate can be close to 0 at the exceptional point. Under the situation of broken PT symmetric Hamiltonian,  we also obtain that non-maximal multiparticle entanglement can improve the precision to reach Heisenberg limit.

The rest of this article is arranged as follows. In Section II, we briefly introduce the quantum Fisher information, and the practical formula of quantum Fisher information for a qubit. In Section III, we detail the PT symmetric non-Hermitian feedback model and show that non-Hermitian feedback can obtain better precision of damping rate than the cases with Hermitian feedback and without any feedback. Then, we obtain the damping parameter precision in an effective PT symmetric Hamiltonian model in Section IV.  A conclusion and outlook are presented in Section V.
\section{review of quantum Fisher information }
The famous Cram$\acute{e}$r-Rao bound\cite{lab31,lab32} offers a very good parameter estimation under the constraints of quantum physics:
\begin{eqnarray}
(\delta x)^2\geq\frac{1}{NF[\hat{\rho}_S(x)]},
\end{eqnarray}
where $N$ represents total number of experiments.  $F[\hat{\rho}_S(x)]$ denotes QFI, which can be generalized from classical Fisher information. The classical Fisher information is defined by
\begin{equation}
f(x)=\sum_k p_k(x)[d\ln[p_k(x)]/dx]^2,
\end{equation}
where $p_k(x)$ is the probability of obtaining the set of experimental results $k$ for the parameter value $x$. Furthermore,
the QFI is given by the maximum of the Fisher information over all measurement strategies allowed by quantum physics:
\begin{equation}
F[\hat{\rho}(x)]=\max_{\{\hat{E}_k\}}f[\hat{\rho}(x);\{\hat{E}_k\}],
\end{equation}
where positive operator-valued measure $\{\hat{E}_k\}$ represents a specific measurement device.

If the probe state is pure, $\hat{\rho}_S(x)=|\psi(x)\rangle\langle\psi(x)|$, the corresponding expression of QFI is
\begin{equation}
F[\hat{\rho}(x)]=4[\frac{d\langle\psi(x)|}{dx}\frac{d|\psi(x)\rangle}{dx}-|\frac{d\langle\psi(x)|}{dx}|\psi(x)\rangle|^2].
\end{equation}
If the probe state is mixed state, $\hat{\rho}(x)=\sum_k\lambda_k|k\rangle\langle k|$, the concrete form of QFI is given by
\begin{equation}
F[\hat{\rho}(x)]=\sum_{k,\lambda_k>0}\frac{(\partial_x \lambda_k)^2}{\lambda_k}+\sum_{k,k',\lambda_k+\lambda_k'>0}\frac{2 (\lambda_k- \lambda_k')^2}{\lambda_k+\lambda_k'}\langle k|\partial_x|k'\rangle.
\end{equation}
In general, it is complicated to calculate QFI. In this paper, we only consider two-dimensional system.  The QFI can be calculated explicitly by the following way\cite{lab33,lab34}
\begin{equation}
F[\hat{\rho}(x)]=\textmd{Tr}[(\partial_x\hat{\rho}(x))^2]+\frac{1}{\textmd{Det}(\hat{\rho}(x))}\textmd{Tr}[(\hat{\rho}(x)\partial_x\hat{\rho}(x))^2].
\end{equation}

\section{PT symmetric non-Hermitian feedback}

\begin{figure}[h]
\includegraphics[scale=0.45]{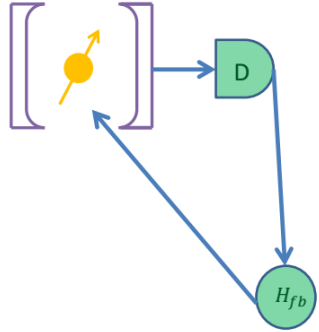}
 \caption{\label{fig.1} Schematic for a photodetection measurement-based feedback control protocol. Information about the system is
extracted by using the detector D to monitor the output from the cavity. Then the signal I(t) from the detector D triggers
the control Hamiltonian $H_{fb} = I(t)B$.}
 \end{figure}
 \begin{figure}[h]
\includegraphics[scale=0.8]{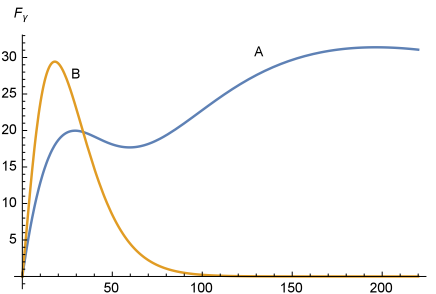}
 \caption{\label{fig.2} The evolution of the QFI $F_\gamma$ with time $t$. The line A represents the case with the unbroken PT symmetric feedback Hamiltonian. The line $B$ denotes the  case without feedback. The
parameters are given by :$a=10\sqrt{2}$, b=10, $\gamma=0.1$. Here, we emphasize that all the relevant quantity is dimensionless in this figure and
 the following other figures.}
 \end{figure}
 \begin{figure}[h]
\includegraphics[scale=0.6]{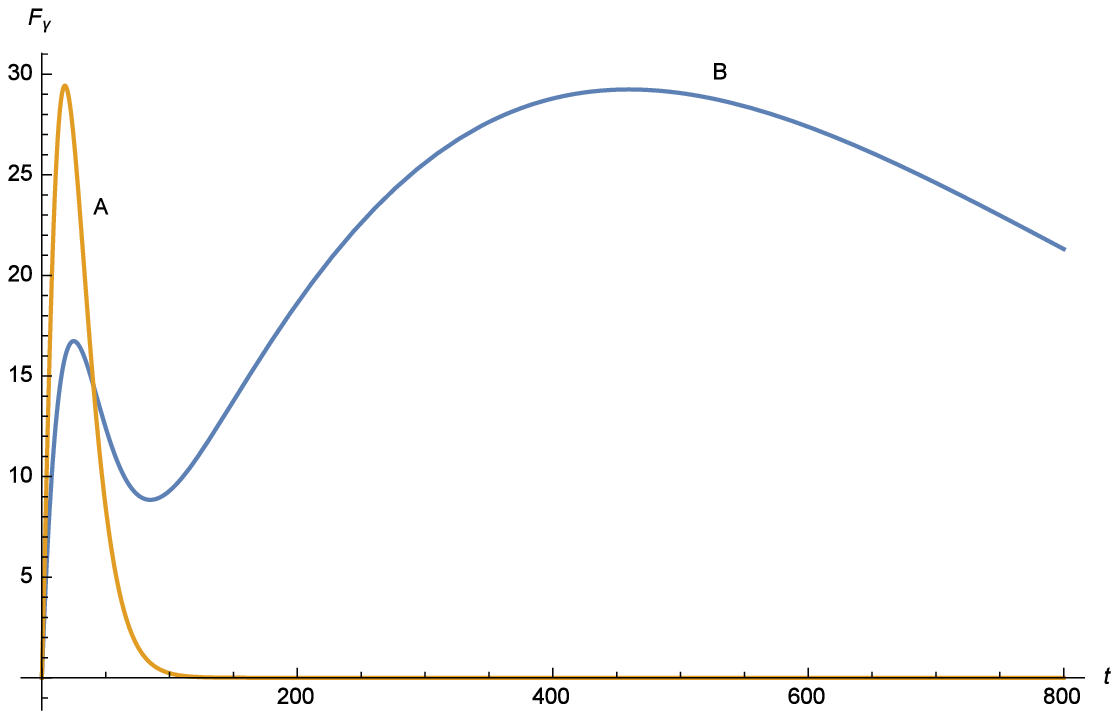}
 \caption{\label{fig.3} The evolution of the QFI $F_\gamma$ with time $t$. The line A represents the case with the PT symmetric feedback Hamiltonian at the exceptional point. The line $B$ denotes the  case without feedback.   The
parameters are given by: $a=0.8$, b=-0.8, $\gamma=0.1$.}
 \end{figure}
  \begin{figure}[h]
\includegraphics[scale=0.8]{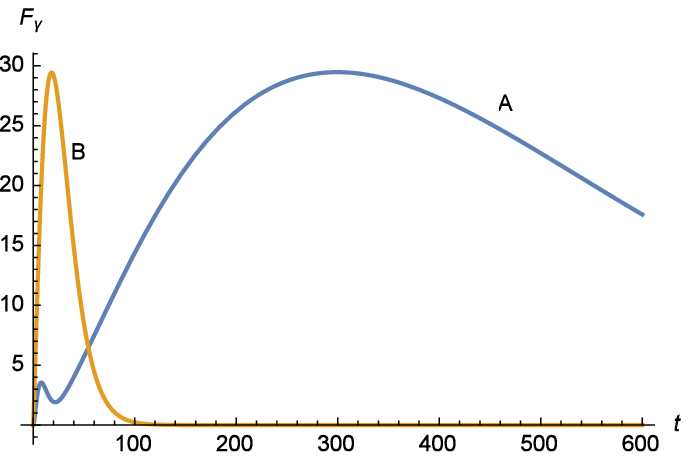}
 \caption{\label{fig.4} The evolution of the QFI $F_\gamma$ with time $t$. The line A represents the case with the broken PT symmetric feedback Hamiltonian. The line $B$ denotes the  case without feedback.  The
parameters are given by: $a=1$, b=-2, $\gamma=0.1$.}
 \end{figure}
   \begin{figure}[h]
\includegraphics[scale=0.7]{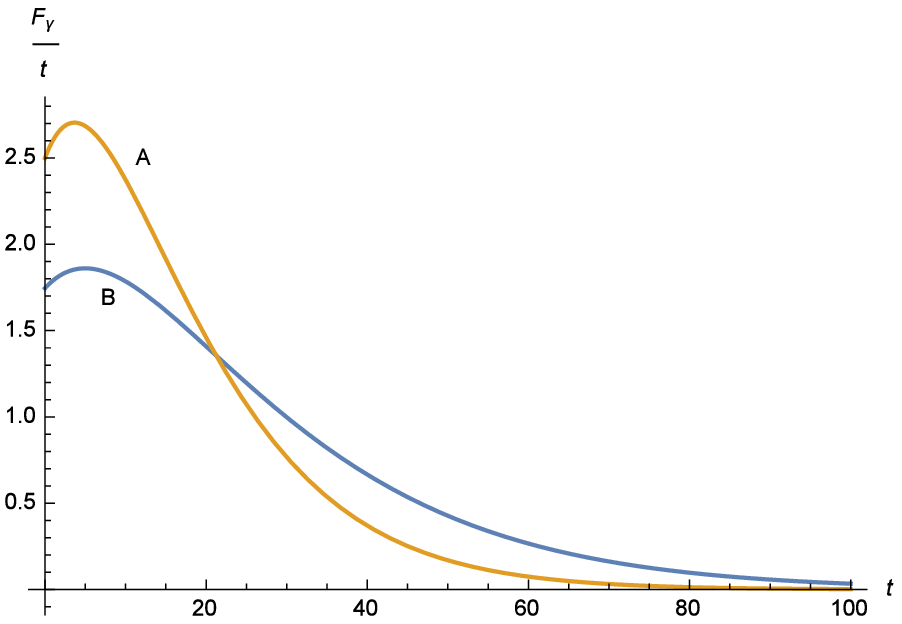}
 \caption{\label{fig.5}The evolution of the $F_\gamma/t$ with time $t$. The line A represents the case with the unbroken PT symmetric feedback Hamiltonian. The line $B$ denotes the  case without feedback.  The
parameters are given by: $a=5$, b=4, $\gamma=0.1$.}
 \end{figure}
  \begin{figure}[h]
\includegraphics[scale=0.7]{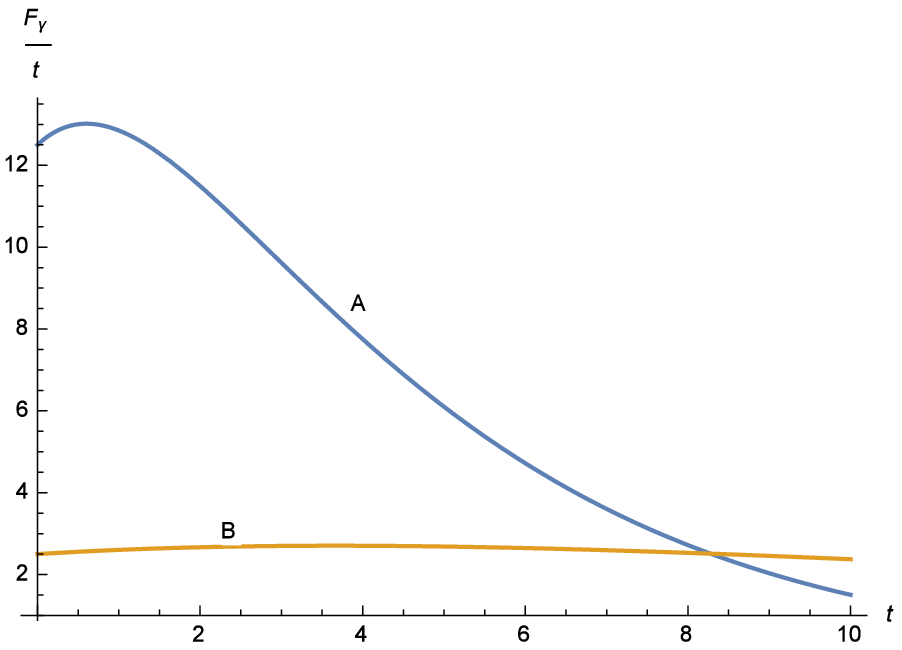}
 \caption{\label{fig.6} The evolution of the $F_\gamma/t$ with time $t$. The line A represents the case with the PT symmetric feedback Hamiltonian at the exceptional point. The line $B$ denotes the  case without feedback.  The
parameters are given by: $a=1$, b=1, $\gamma=0.1$.}
 \end{figure}
  \begin{figure}[h]
\includegraphics[scale=0.7]{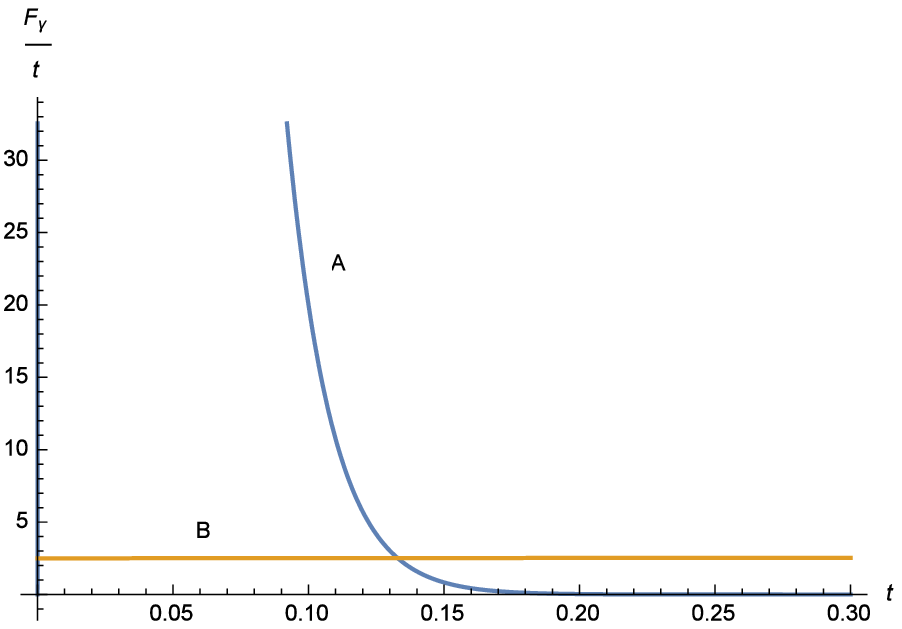}
 \caption{\label{fig.7}The evolution of the $F_\gamma/t$ with time $t$. The line A represents the case with the broken PT symmetric feedback Hamiltonian. The line $B$ denotes the  case without feedback.  The
parameters are given by :$a=4$, b=5, $\gamma=0.1$.}
 \end{figure}
We consider  a two-level system($|e\rangle$, $|g\rangle$), which resonantly interacts with a single-mode cavity, as shown in Fig. 1. Without feedback, the master equation of system can be described by ($\hbar=1$ throughout this article)
\begin{equation}
\frac{d\rho}{dt}=-i[H,\rho]+\gamma \mathcal{D}[\sigma_- ]\rho,
\end{equation}
where the Pauli operator is described by $\sigma_-=|g\rangle\langle e|$, the superoperator  is defined as $\mathcal{D}[c ]\rho=c\rho c^\dagger-\frac{1}{2}(c^\dagger c\rho+\rho c^\dagger c)$ and $\gamma$ is the damping rate.

We consider a feedback as shown in Fig.1: the feedback Hamiltonian $H_{fb}=I(t)B$, where the signal $I(t)$ is obtained from the detector $D$ by the direct photodetection measurement. The unconditional master equation of the system is described by \cite{lab24,lab35,lab36}
\begin{equation}
\frac{d\rho}{dt}=-i[H,\rho]+\gamma \mathcal{D}[U\sigma_- ]\rho,
\end{equation}
where $\mathcal{D}[U\sigma_- ]\rho=U\sigma_- \rho\sigma_+U^\dagger-\frac{1}{2}(\sigma_+U^\dagger U\sigma_-\rho+\rho\sigma_+U^\dagger U\sigma_-)$ and $H=\Omega \sigma_x $. In this article, we consider that the feedback Hamiltonian is non-Hermitian, $B\neq B^\dagger$. Therefore, the transformation operator $U=\exp[-iB\delta t]$, which is not unitary evolution. Without loss of generality, we set $ \delta t=1$ throughout this article.
In ref.\cite{lab37}, a optimal feedback operator is chosen as $B=a\sigma_x+b\sigma_y$ with a (b) denotes a real number. In ref.\cite{lab24}, $B=a\sigma_x$ is shown to be a good approximation. In this article, we consider the PT symmetric non-Hermitian feedback operator $B=a\sigma_x+i b \sigma_z$. This minimal model has been studied by a lot of works\cite{lab38,lab39,lab40,lab41}. When $a^2>b^2$, it is unbroken PT symmetric non-Hermitian Hamiltonian; when $a^2<b^2$, it is broken PT symmetric non-Hermitian Hamiltonian; when $a^2=b^2$, the Hamiltonian is at exceptional point\cite{lab42}.

For a superposition initial state $|\psi(0)\rangle=\frac{1}{\sqrt{2}}(|e\rangle+|g\rangle)$ and without external driving ($\Omega=0$), the evolved density matrix of the qubit can be exactly solved, which is given as
 \[
 \rho(t)= \left(
\begin{array}{ll}
\rho_{11}(t)\ \ \rho_{12}(t)\\
\rho_{12}^*(t)\ \ 1-\rho_{11}(t)
  \end{array}
\right ),
  \]
 in which,

   \[
\textmd{for unbroken PT}\  (a^2>b^2),  \left\{
\begin{array}{ll}
\rho_{11}(t)=1-\frac{1}{2}\exp[-\gamma t(\cos{q}+\frac{b}{q}\sin q)^2];\\
\rho_{12}(t)=-\Gamma\exp[-\gamma t(\cos{q}+\frac{b}{q}\sin{q})^2]+(\frac{1}{2}+\Gamma)\exp[-\frac{\gamma}{2} t((\cos{q}+\frac{b}{q}\sin{q})^2+(\frac{a}{q}\sin{q})^2)];\\
q=\sqrt{a^2-b^2}, \Gamma=\frac{i(\cos{q}+\frac{b}{q}\sin q)\frac{a}{q}\sin{q}}{(\frac{a}{q}\sin{q})^2-(\cos{q}+\frac{b}{q}\sin q)^2},
  \end{array}
  \right.
  \]\begin{equation}
 \end{equation}
    \[
\textmd{for exceptional point}\   (a^2=b^2), \left\{
\begin{array}{ll}
\rho_{11}(t)=1-\frac{1}{2}\exp[-\gamma t(1+b)^2];\\
\rho_{12}(t)=\frac{i(1+b)b}{2b+1}\exp[-\gamma t(1+b)^2]+(\frac{1}{2}-\frac{i(1+b)b}{2b+1}\exp[-\frac{\gamma }{2} t((1+b)^2+b^2)],
  \end{array}
  \right.
  \]\begin{equation}
 \end{equation}
   \[
\textmd{for broken PT}\   (a^2<b^2),  \left\{
\begin{array}{ll}
\rho_{11}(t)=1-\frac{1}{2}\exp[-\gamma t(\cos{i q}-\frac{i b}{q}\sin i q)^2];\\
\rho_{12}(t)=-\Gamma\exp[-\gamma t(\cos{i q}-\frac{i b}{q}\sin{i q})^2]+(\frac{1}{2}+\Gamma)\exp[-\frac{\gamma}{2} t((\cos{i q}-\frac{i b}{q}\sin{i q})^2+(\frac{i a}{q}\sin{i q})^2)];\\
q=\sqrt{a^2-b^2}, \Gamma=\frac{i[\cos{i q}-\frac{i b}{q}\sin i q]\frac{a}{q}\sin{i q}}{[\frac{i a}{q}\sin{i q}]^2-[\cos{i q}-\frac{i b}{q}\sin i q]^2}.
  \end{array}
  \right.
  \]\begin{equation}
 \end{equation}

The optimal precision of the damping parameter $\gamma$ can be obtained by Eq.(6). However, the general analytical expression is very cumbersome. We can obtain the numerical result as shown in Fig.2-7. Fig.2-4 represent the optimal quantum Fisher information of damping parameter $\gamma$ under the above three cases: unbroken PT symmetric feedback Hamiltonian(Fig.2), exceptional point(Fig.3) and broken PT symmetric feedback Hamiltonian(Fig.4).
Compared with the line B (without feedback Hamiltonian), we can find a marked difference that the line A (with non-Hermitian feedback) has two peaks. There is a peak in the case of Hermitian feedback Hamiltonian\cite{lab24}. It can be attributed to non-Markovianity from the non-Hermitian feedback. The information can return from the environment to the system\cite{lab43}. However, in ref.\cite{lab43}, the information can turn back only in the unbroken PT symmetric feedback. In our feedback model, we find that the quantum Fisher information can increase again (meaning information backflow from the environment) in the cases of broken PT symmetric feedback and exceptional point.
 It merits further study of the essence.

 From Fig.2, we can see that the non-Hermitian feedback can obtain greater QFI than the case without feedback. And from Fig.2-4, the QFI with the feedback decays more slowly
than that without the feedback in the long time. Hence, the results are similar with the result from Hermitian feedback, as shown in ref. \cite{lab24}. We can also obtain new results: by choosing different parameters, the QFI with the non-Hermitian feedback can also decays more quickly than that without the feedback. Given the fixed measurement time $T$, the precision of damping parameter $\gamma$ can be described by
\begin{eqnarray}
(\delta \gamma)^2\geq\frac{1}{\frac{T}{t}F[\hat{\rho}_S(\gamma)]}.
\end{eqnarray}
So the higher precision of damping rate $\gamma$, the larger value of $F/t$. It can be shown in Fig.5-7, the QFI with the non-Hermitian feedback can be larger than that without feedback at short time.
In order to better understand the numerical result, we can consider a result obtained with a fixed projective measurement, which is given by the measurement operator ($|e\rangle\langle e|$, $|g\rangle\langle g|$).
With the projective measurement, the Fisher information is calculated by Eq.(2)
\begin{eqnarray}
f=\frac{\exp[-\gamma t(\cos{q}+\frac{b}{q}\sin{q})^2]t^2(\cos{q}+\frac{b}{q}\sin{q})^4}{2-\exp[-\gamma t(\cos{q}+\frac{b}{q}\sin{q})^2]},
\end{eqnarray}
where $q=\sqrt{a^2-b^2}$.
From this equation, we can see that the feedback factor $(\cos{q}+\frac{b}{q}\sin{q})^2$ can influence the damping rate $\gamma$. For the case of Hermitian feedback ($b=0$), the feedback factor $(\cos{q}+\frac{b}{q}\sin{q})^2<1$ . However, for  the case of Hermitian feedback ($b\neq0$), the factor can be larger than 1 so that the decay rate is increased. The maximal value of $f/t$ can be obtained approximately at $t=\frac{1}{\gamma(\cos{q}+\frac{b}{q}\sin{q})^2}$,
\begin{eqnarray}
\frac{f}{t}|_{M}\approx\frac{ (\cos{q}+\frac{b}{q}\sin{q})^2}{(2e-1)\gamma}.
\end{eqnarray}
Therefore, the optimal precision of damping rate $\gamma$ can be enhanced by increasing the feedback factor $(\cos{q}+\frac{b}{q}\sin{q})^2$  under the situation of considering the resource of time.
This can help us to understand the result shown in Fig.5-7.

 \section{The damping parameter encoded in an effective PT symmetric Hamiltonian model}
 Conditioned on the absence of decay events\cite{lab44,lab45}, the term $\sigma_-\rho\sigma_+$ in Eq.(7) can be removed. As a result, the second term of Eq.(7) is written as $-\gamma(|e\rangle\langle e|\rho+\rho|e\rangle\langle e|)$.
The corresponding conditional master equation is described by $\frac{\partial \rho}{\partial t}=-i(H_{eff}\rho-\rho H_{eff}^\dagger )$, where $H_{eff}$ is the effective PT symmetric Hamiltonian,
\begin{eqnarray}
H_{eff}=\Omega\sigma_x-i\gamma|e\rangle\langle e|.
\end{eqnarray}

Firstly, we consider that there is no external driving ($\Omega=0$). We utilize the entangled state $\cos(\theta)|e\rangle^{\otimes N}+\sin(\theta)|g\rangle^{\otimes N}$ of the same $N$ systems to improve the precision of $\gamma$. Normalized density matrix $\varrho(t)$ is given by\cite{lab46}
\begin{eqnarray}
\varrho(t)=\frac{(\exp[-i H_{eff} t]\cos\theta|e\rangle^{\otimes N}+\sin\theta|g\rangle^{\otimes N})(\cos\theta\langle e|^{\otimes N}+\sin\theta\langle g|^{\otimes N}\exp[i H_{eff}^\dagger t])}{\textmd{Tr}[(\exp[-i H_{eff} t]\cos\theta|e\rangle^{\otimes N}+\sin\theta|g\rangle^{\otimes N})(\cos\theta\langle e|^{\otimes N}+\sin\theta\langle g|^{\otimes N}\exp[i H_{eff}^\dagger t])]}\\
=\frac{(\cos\theta\exp[-\gamma Nt]|e\rangle^{\otimes N}+\sin\theta|g\rangle^{\otimes N})(\exp[-\gamma Nt]\cos\theta\langle e|^{\otimes N}+\sin\theta\langle g|^{\otimes N})}{\cos^2\theta\exp[-2\gamma Nt]+\sin^2\theta}.
\end{eqnarray}
 Substituting Eq.(17) into Eq.(6), we can obtain the analytical expression of QFI
 \begin{eqnarray}
F_\gamma=\frac{2\cos^2 \theta\sin^2 \theta \exp[2\gamma Nt ](Nt)^2}{(\cos^2 \theta+\sin^2 \theta\exp[2\gamma Nt ])^2}.
\end{eqnarray}
As a result, the corresponding precision of damping rate $\gamma$ is given by
\begin{eqnarray}
(\delta \gamma)^2\geq\frac{(\cos^2 \theta+\sin^2 \theta\exp[2\gamma Nt ])^2T}{2\cos^2 \theta\sin^2 \theta \exp[2\gamma Nt ]N^2t},
\end{eqnarray}
where $T$ denotes the given total interrogation time.
From the above Eq.(19), we can obtain that for the maximally entangled state ($\sin^2\theta=1/2$) and $N\gg1$, the optimal precision is proportional to $1/N$, which is called the quantum limit.

When we measure the $N$ systems at time $t=1$ with the initial parameter $\sin^2\theta=\exp[-2 \gamma N]$, the optimal precision of damping rate $\gamma$ is obtained
\begin{eqnarray}
(\delta \gamma)^2=\frac{1}{2N^2}.
\end{eqnarray}
Namely, Heisenberg limit of damping rate $\gamma$ has been achieved by using a small entangled state $\cos\theta|e\rangle^{\otimes N}+\sin\theta|g\rangle^{\otimes N}$. In one word, non-maximally entangled state can help to achieve a better precision of damping rate than that with the maximally entangled state.

Then, we consider that there is an external driving  ($\Omega\neq0$). We use the eigenstate of the non-Hermitian Hamiltonian $H_{eff}=\Omega\sigma_x-i\gamma|e\rangle\langle e|$ to measure the parameter $\gamma$. The corresponding two non-normalized eigenstates are described by
 \[
 |\Psi_-\rangle= \left(
\begin{array}{ll}
\ -\Omega\\
i \gamma-\sqrt{\Omega^2-\gamma^2}
  \end{array}
\right ), |\Psi_+\rangle= \left(
\begin{array}{ll}
\ -\Omega\\
i \gamma+\sqrt{\Omega^2-\gamma^2}
  \end{array}
\right ).
  \]
Normalizing the above eigenstates and utilizing Eq.(4), we can obtain the QFI of damping rate with $\Omega\geq\gamma$
\begin{eqnarray}
F_\gamma=\frac{2}{\Omega^2-\gamma^2}.
\end{eqnarray}
Therefore, we find that at the exceptional point ($\Omega=\gamma$), the QFI becomes infinity. Namely, one can utilize the exceptional point to obtain a very perfect precision of damping rate $\gamma$: $\delta\gamma=0$.
\section{conclusion and outlook}
We have utilized two different models to measure the damping rate $\gamma$ and obtain the corresponding precision. The results show that direct PT symmetric non-Hermitian quantum feedback can obtain a larger QFI of the damping rate $\gamma$ than that without quantum feedback. When considering the resource of interrogation time,  direct PT symmetric non-Hermitian quantum feedback can obtain a better precision than the case with Hermitian feedback. It is due to that non-Hermitian quantum feedback can make the feedback factor be larger than 1. When the damping rate is encoded into an effective PT symmetric Hamiltonian, we achieve that using a small entangled state can help to enhance the precision of parameter to reach Heisenberg limit. And we find that the uncertainty of damping rate can be 0 at the exceptional point.

Our results show that PT symmetric Hamiltonian can help to obtain better precision of damping rate. It will motivate the further study of PT symmetric Hamiltonian in quantum metrology.

\section*{Acknowledgement}
 This research was supported by the National
Natural Science Foundation of China under Grant No. 11747008 and Guangxi Natural Science Foundation 2016GXNSFBA380227.

\end{document}